\documentclass[]{aa}

\usepackage{graphicx,epsfig}
\usepackage{natbib}
\usepackage[usenames]{color}
\usepackage{txfonts}
\usepackage{url}
\usepackage[dvips]{hyperref}
\usepackage{stfloats}

\newcommand{\degree}{\ensuremath{^\circ}}
\newcommand{\hinode}{{\em Hinode}}
\newcommand{\soho}{{\em SOHO}}
\newcommand{\stereo}{{\em STEREO}}
\newcommand{\sta}{{\em STEREO-A}}
\newcommand{\stb}{{\em STEREO-B}}
\newcommand{\pref}{\protect\ref}

\begin{document}

\title{STEREO Observations of Quasi-Periodically Driven High Velocity Outflows in Polar Plumes}
\author{Scott W. McIntosh\inst{1}, Davina E. Innes\inst{2}, Bart De Pontieu\inst{3}, and Robert J. Leamon\inst{4,5}}
\institute{High Altitude Observatory, National Center for Atmospheric Research, P.O. Box 3000, Boulder, CO 80307, USA \and Max-Planck Institut f\"{u}r Sonnensystemforschung, 37191 Katlenburg-Lindau, Germany \and Lockheed Martin Solar and Astrophysics Lab, 3251 Hanover St., Org. ADBS, Bldg. 252, Palo Alto, CA  94304, USA \and ADNET Systems Inc., NASA/GSFC, Greenbelt, MD 20771, USA \and now at Department of Physics, Montana State University, Bozeman, MT 59717, USA}
\offprints{S.~W. McIntosh \email{mscott@ucar.edu}}
\date{Received ... / Accepted ...}
\abstract
{Plumes are one of the most ubiquitous features seen at the limb in polar coronal holes and are considered to be a source of high density plasma streams to the fast solar wind.}
{We analyze \stereo{} observations of plumes and aim to reinterpret and place observations with previous generations of EUV imagers within a new context that was recently developed from \hinode{} observations.}
{We exploit the higher signal-to-noise, spatial and temporal resolution of the EUVI telescopes over that of \soho/EIT to study the temporal variation of polar plumes in high detail. We employ recently developed insight from imaging (and spectral) diagnostics of active region, plage, and quiet Sun plasmas to identify the presence of apparent motions as high-speed upflows in magnetic regions as opposed to previous interpretations of propagating waves.}
{In almost all polar plumes observed at the limb in these \stereo{} sequences, in all coronal passbands, we observe high speed jets of plasma traveling along the structures with a mean velocity of 135~km/s at a range of temperatures from 0.5-1.5~MK. The jets have an apparent brightness enhancement of $\sim$5\% above that of the plumes they travel on and repeat quasi-periodically, with repeat-times ranging from five to twenty-five minutes. We also notice a very weak, fine scale, rapidly evolving, but ubiquitous companion of the plumes that covers the entire coronal hole limb.}
{The observed jets are remarkably similar in intensity enhancement, periodicity and velocity to those observed in other magnetic regions of the solar atmosphere. They are multi-thermal in nature. We infer that the jets observed on the plumes are a source of heated mass to the fast solar wind. Further, based on the previous results that motivated this study, we suggest that these jets originated in the upper chromosphere.}


\titlerunning{Ubiquitous Jets on Polar Plumes}
\authorrunning{McIntosh et~al.}
\maketitle

\section{Introduction}
Plumes are one of the most ubiquitous features seen at the limb in polar coronal holes \citep[e.g.,][]{Newkirk1968}. In recent years they have been studied at great detail \citep[see, e.g.,][as a representative few]{Deforest1997, Ofman1999, Banerjee2009}, are thought of sources of dense plasma in the fast solar wind \citep[e.g.,][]{Gabriel2003} that result from the relentless magnetoconvective forcing of the upper solar atmospheric plasma \citep[e.g.,][]{Wang1998, McIntosh2007}.

\begin{figure}
\resizebox{\hsize}{!}{\includegraphics{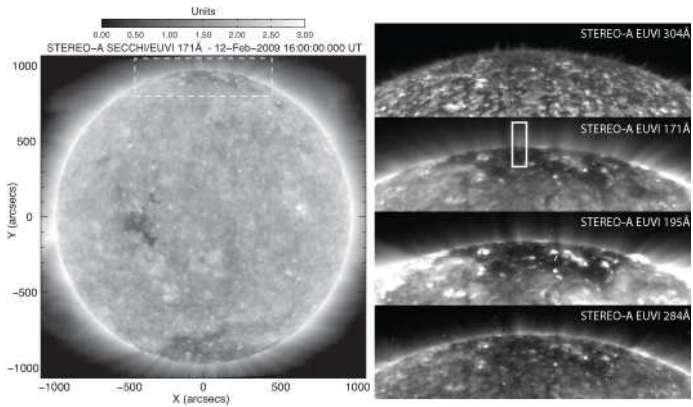}}
\caption{Context of the observations provided by the \sta{} SECCHI/EUVI 171\AA{} passband. The North PCH region that provides the focus of the presented analysis is shown in the box. The inset panels on the right of the figure show the first 304\AA{}, 171\AA{}, 195\AA{}, and 284\AA{} images of the February 12 2009 sequence. Also shown in the inset panels are the location of the synthetic slit used to generate the timeseries shown in Fig.~\pref{f2}. The online edition of the journal has movies in support of the four inset panels.} \label{f1}
\end{figure}

In this Letter we use high cadence observations by the \stereo{} spacecraft to explore the weak, fine, temporal variations of the emitting material that appear to be ubiquitous in polar plumes. We use a simple space-time data analysis technique similar to that used in the investigation of quasi-periodic intensity perturbations in active regions. \citet{McIntosh2009a} followed from \citet{DePontieu2009} to demonstrate that these weak, quasi-periodic, perturbations in coronal imaging data are directly related with the weak, high speed Doppler velocity signals in the blue wing of several coronal emission lines. Further, these blue-wing asymmetries are co-spatial and co-temporal with high speed chromospheric jets, or ``Type-II'' spicules \citep[][]{DePontieu2007}, rooted in the magnetic footpoints of the active region. \citet{McIntosh2009b} \citep[and][]{McIntosh2010} showed that these upflows are ubiquitous also in magnetic network regions (of the quiet sun and coronal holes), again occurring quasi-periodically with repeat times ranging from a few, to several tens of minutes, and with speeds determined from the analysis of line profile asymmetries of many spectral lines (formed over a broad range of formation temperatures) that are similar to those observed in the apparent motions derived from coronal imaging data \citep[e.g.,][]{Schrijver1999}.

In the following sections we discuss the \stereo{} observations used, and the analysis techniques employed. Based on an argument of similarity with spectroscopic and imaging features observed ubiquitously in quiet, coronal hole, and active solar plasmas alike we suggest that the quasi-periodic jets in polar plumes observed at the limb by \stereo{} are likely to be similarly excited, with roots in dynamic upper chromospheric activity. We conclude by discussing the implications of this new interpretation, looking forward to observational tests of the paradigm and closing with speculation on the impact of our interpretation on the process of solar wind acceleration.

\section{Observations \& Reduction}
The observations discussed in this Letter were taken by the EUVI/SECCHI telescopes on the \stereo{} spacecraft from 14:00-18:00~UT on 12-15 Feb 2009. During the four hours of observation on each day, \stereo{}-A took 171\AA{} passband images every 75s that were synchronized with 304\AA{} images of \stb. The EUVI telescopes also took a slightly lower cadence (95s) synchronized sequence of  \sta{} 304\AA{} and \stb{} 171\AA{} over the same period with context images in the 195\AA{} and 284\AA{} passbands every 10 minutes. At this time \stb{} was 47.5\degree\ behind the earth and \sta was 43.4\degree\ ahead, yielding a total spacecraft separation angle of 91\degree.

The EUVI/SECCHI data have been calibrated using the standard SolarSoft IDL routines. They were corrected for the telescope roll angle and deprojected so that the solar B angle is zero, for all images. The disk center images were then corrected for differential rotation, and aligned to the start time of each sequence. We then expanded the Sun in the B images to match the radius of the A images. After using the numbers from the headers, we still found small corrections were necessary to achieve perfect co-alignment of features. The \sta{} images required an additional rotation of 2\degree\ and the Sun center in the \stb{} images was shifted north by 5\arcsec.

\section{Analysis \& Results}
We have isolated the North polar coronal hole (PCH) region in the EUV imaging sequences shown in Fig.~\pref{f1}. The online edition of the journal contains movies of the 304, 171, 195, and 284\AA{} passbands of the A spacecraft in this region. From those movies the temporal variability of the polar plume above the north polar limb ($x=-22\arcsec$) is easily observed in all three coronal passbands. Connection of the plume to the low transition region plasma (304\AA) is complicated by the prominent fore- and background spicules (and macrospicules). We interpret the very weak emission seen in the 284\AA{} passband as that of the \ion{Si}{VII} 275\AA{} emission line that is formed, in equilibrium, at 0.6MK \citep{Mazzotta1998} and dominates the passband in quieter network regions near solar minimum.

\begin{figure}
\resizebox{\hsize}{!}{\includegraphics{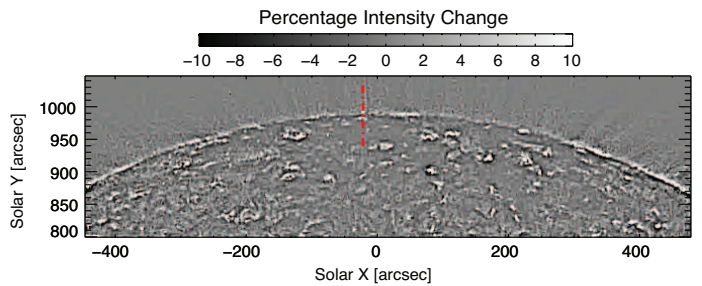}}
\caption{Example frame from the background-subtracted \sta{} 171\AA{} sequence. The synthetic slit used to compute the timeseries shown in Fig.~\pref{f3} is shown as a red dot-dashed line. The online edition of the journal has a movie in support of this figure.} \label{f2}
\end{figure}

To quantify the variability in the plume we produce a timeseries of background subtracted images by removing a 5pixel (7.75\arcsec{}) boxcar smoothed version of each image in the 171\AA{} sequence from itself. As a result of this processing only the finer spatial scale fluctuations remain \citep[cf.][]{McIntosh2009a}. Placing a 100\arcsec{} long ``slit'' at $-22\arcsec, 940\arcsec{}$ normal to the limb for the 171\AA{} \sta{} sequence we obtain the timeseries shown in Fig.~\pref{f2}.

\begin{figure}
\resizebox{\hsize}{!}{\includegraphics{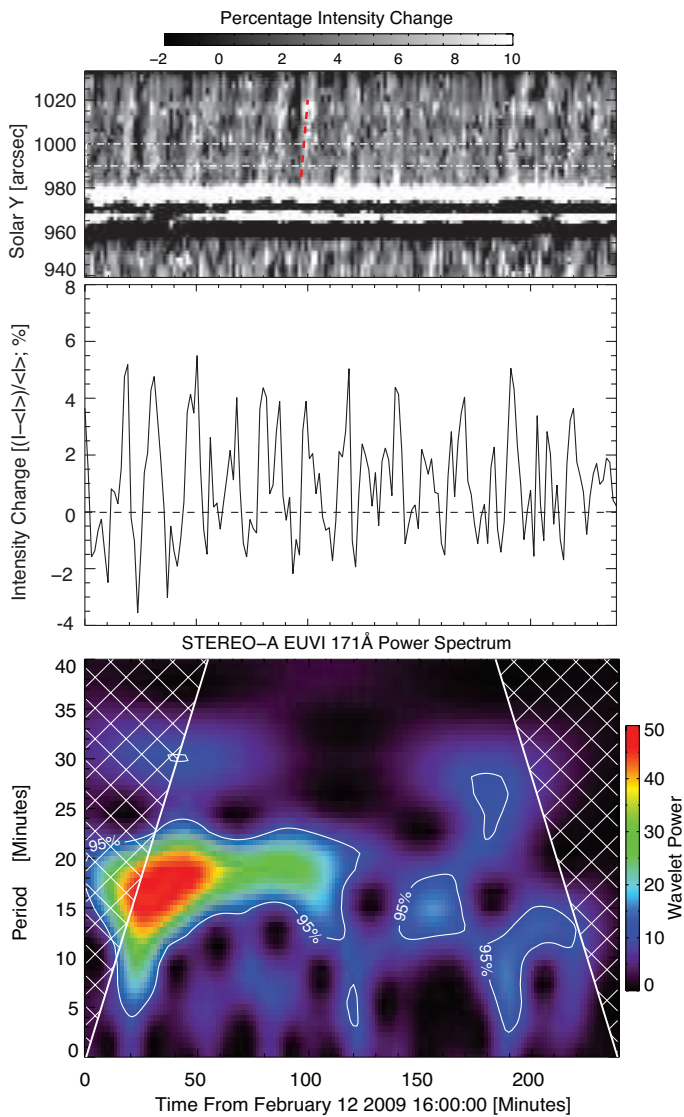}}
\caption{Timeseries analysis of a plume in the \sta{} 171\AA{} image sequence derived from the synthetic slit position shown in Fig.~\pref{f2}. From top to bottom we see the percentage change in the background-subtracted timeseries imaged along the slit, a spatially-averaged cut through the slit timeseries (dot-dashed region in the top panel), and the wavelet power spectrum and its 95\% confidence level (white contours).} \label{f3}
\end{figure}

In Fig.~\pref{f3} we assess the temporal variation of the emission along the slit. In the top panel we show the percentage change in the background-subtracted timeseries imaged along the slit and see the series of inclined alternating black and white stripes. The white stripes represent intensity enhancements of $\sim$5-7\%. The inclination of the stripes indicates the propagation speed of the disturbance along the chosen trajectory where, for reference, we show a red line on the jet starting 95 minutes into the timeseries with a velocity of 142 \citep[$\pm$ 10~km/s, cf.][]{Deforest1998, Lites1999}. This velocity is representative of the complete sample as there is only 15~km/s variance throughout the four hours of observation indicating that the apparent motion is a function of the plume's thermodynamic or magnetic properties. The occurrence of these disturbances can be determined by taking a spatially-averaged cut through the slit timeseries ($y = 990-1000\arcsec$; dot-dashed region in the top panel) and shown in the center panel. We see that the occurrence of the disturbances is roughly quasi-periodic with events occurring anywhere from 5 to $\sim$15 minutes apart. Their strength or amplitude shows less variability ($\sim$6\% enhancement) as we have noted above. The quasi-periodic nature of the occurrence of these disturbances can be viewed in another way by computing the wavelet power spectrum of the timeseries as is shown in the bottom panel of Fig.~\pref{f3}. The regions of the wavelet power spectrum enclosed in contours are those which have 95\% statistical significance \citep[][]{Torrence1998}. The wavelet power spectrum confirms our visual inspection of the spatial cut through the timeseries --- it shows a broad range of significant periods but one of predominantly 15-18 minutes ($\sim$1mHz) for a large part of timeseries. Such quasi-periodicities have been noted before by \citet{Deforest1998} and \citet{Banerjee2000, Banerjee2009} where they were interpreted as belonging to propagating slow-mode magneto-acoustic waves.

\begin{figure}
\begin{center}
\resizebox{6cm}{!}{\includegraphics{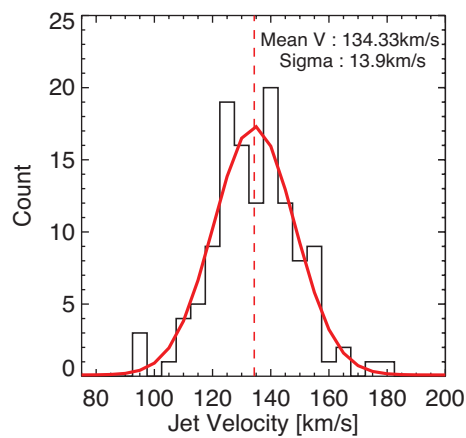}}
\end{center}
\caption{Distribution of jet velocities clearly visible in ten plumes characterized. The sample includes information derived from 123 individual jets (see, eg., Fig.~\pref{f3}).} \label{f4}
\end{figure}

Using a tool like the {\tt xslice} routine (publically available in the SolarSoft IDL analysis tree) permits the exploration of the 171\AA{} image sequences from both spacecraft, in both polar regions, and to characterize the appearance of ten clearly defined plumes present in the data. Using the method of \citet{McIntosh2004} to determine the plume repeat times we characterize the wavelet power spectra we see plume disturbance periods in a range of 16.6 ($\pm$ 6.6) minutes. For the ten plumes studied in detail we can easily discern 123 individual events and Fig.~\pref{f4} shows the derived apparent motions. The distribution of their velocities has a mean of 134 ($\pm$ 14) km/s where the width of the distribution incorporates likely error in deducing the gradients in the x-t plots.

Finally, very close inspection of the movie supporting Fig.~\pref{f2} suggest the presence of a ubiquitous, very weak, fine scale, rapidly evolving, ``mist'' of upward propagating companions to the plumes that cover the entirety of the coronal hole limb. Unfortunately, the spatial resolution and compression of the \stereo{} spacecraft limit any precise diagnosis of this signal and we must wait for the higher signal-to-noise and spatial resolution images of the Solar Dynamics Observatory's (SDO) Atmospheric Imaging Assembly (AIA) for further insight.

\section{Discussion}
We have observed weak, quasi-periodic, high-speed perturbations in several polar plumes observed by the \stereo{} spacecraft at range of temperatures from 0.6\--1.5MK. These perturbations have very similar properties to those observed in active regions \citep{McIntosh2009a}, coronal holes \citep{DePontieu2009,McIntosh2010}, and quiet Sun \citep{McIntosh2009b}. In each case, these perturbations have been connected spectroscopically to a strong upflowing, weak emission component at the magnetic footpoints. The spectroscopically determined upflows appear to be rooted in dynamic ``Type-II'' spicules in the upper chromosphere \citep{DePontieu2009, McIntosh2009a}. The overwhelming similarity between these results leads us to propose that the quasi-periodically forced jets observed in polar plumes have a similar cause and, as such, are responsible for loading a significant amount of heated ($\ge$1MK) plasma into the fast solar wind along the open magnetic field lines \citep[][]{Parker1991}, a hypothesis supported by high latitude Ulysses/SWOOPS observations \citep[][]{Yamauchi2003}. 

We observe the same types of jets for a wide range of temperatures, from 0.6 to 1.5 MK. Because the \stereo{} data are not taken exactly simultaneously, is not clear whether cool and hot jets co-occur within the same pixel. However, the analysis of \citet{McIntosh2009a} of co-located upflows for a wide range of temperatures, as well as observations of large-scale jets in coronal holes that appear in simultaneous observations at 140,000 and 630,000K \citep{Scullion2009} suggest that these events may well be truly multi-thermal. Observations with SDO/AIA will shed light on this issue, one that has important implications for the driving mechanism of these jets although we expect that they are triggered by the complex interactions of the small scale magnetic flux elements at the base of the plume and its immediate surroundings \citep[][]{Gabriel2009, Heggland2009}.

It is only appropriate to note that our interpretation in terms of field-directed, high-speed, quasi-periodically triggered, upflows is contrary to the widely held interpretation that this observational phenomena is due to compressive (slow-mode) magneto-acoustic waves traveling along the plumes \citep[e.g.,][]{Deforest1998, Ofman1999, Ofman2000, Banerjee2000, Nakariakov2006, Banerjee2009}. Unfortunately, the problem is ill-posed, and both interpretations are limited by the high inclination angles that polar plumes present relative to our typical line of sight on the Sun-Earth line. Further complications arise from: the commonality of the chromospheric Alfv\'{e}n and coronal sound speeds ($\sim$100~km/s); the weak density enhancements due to the upflow, or wave, passage that are largely lost in the lower part of the plume against the emission from the brighter core of the line, only becoming more prevalent as the background emission drops away exponentially with radial distance \citep[][]{Ofman1999}; and issues persist as to why these objects can have such long ``periods'' relative to the typical timescales in the lower atmosphere of only a few minutes if the waves come from, or transit through, the lower atmosphere \citep[see, e.g.,][]{McIntosh2008,Wang2009}. We suggest that efforts should be made to identify plumes in equatorial coronal holes where they can be studied spectroscopically, down their long axis, in order to check the properties of the line emission with the highest possible signal-to-noise. Such observations are planned using detailed \hinode{} SOT and EIS observations (using very deep exposures of the coolest lines in its wavelength range) with corresponding \stereo{} observations at the limb necessary to complete the connection of the plasma from the chromospheric to coronal domain as well as address the wave/upflow interpretation issue. We will investigate the temporal variability in the line core intensity, Doppler shift and broadening down the axis of the plume \citep[studying the phase relationships between them as a wave diagnostic, e.g.,][]{Lites1979}, and check for the presence of blue-wing asymmetries \citep[consistent with the presence of a spicule-related upflows][]{DePontieu2009}. Of course, on disk, we will need to exploit new upper chromospheric diagnostics from SOT analyze the dynamic behavior \citep[e.g.,][]{Rouppe2009}.

Should the observed phenomena be related to quasi-periodic mass-loading events it is natural to expect that the change in tension on that magnetic field line will also trigger a quasi-periodic Alfv\'{e}n wave in the plume \citep[see also][]{Lites1999}. This speculation is supported by the observational evidence of the ubiquitous (and significant) Alfv\'en wave flux carried by the Type-II spicules \citep{DePontieu2007b}. 
Finally, we speculate that because the (Alfv\'{e}n) phase speed of the waves is significantly higher than the field-aligned upflows, the mass on the field line can lead to ideal reflecting conditions for the formation of a turbulent cascade of the Alfv\'enic wave energy into the plasma that is needed to accelerate the wind \citep[e.g.,][]{Velli1993, Matthaeus1999, Verdini2010}. It is likely that precise spectroscopic imaging experiments \citep[e.g.,][]{Tomczyk2007} must be made in polar regions to accurately investigate the propagation and magnitude of Alfv\'{e}n waves in the plume and inter-plume regions.

\section{Conclusion}
We have studied several high cadence \stereo{} sequences of polar coronal holes at the limb. In all coronal passbands, we observe high speed jets of plasma traveling along the plume structures with a mean velocity of $\sim$135~km/s at a range of temperatures from 0.5-1.5~MK. These jets have an apparent brightness enhancement of $\sim$5\% above that of the plumes they travel on and repeat quasi-periodically with repeat-times ranging from five to 25 minutes. These jets are remarkably similar in magnitude, periodicity and velocity to those observed in other magnetic regions of the solar atmosphere. They are multi-thermal in nature. Further, based on the previous results that motivated this study \citep{DePontieu2009,McIntosh2009a,McIntosh2009b}, we speculate that these jets originate in the upper chromosphere and that their quasi-periodic nature can be a ready source of Alfv\'{e}n waves and mass which can self-consistently form and accelerate the mass in the fast solar wind.

\begin{acknowledgements}
We have benefited and are grateful for frequent discussions with Joe Gurman and Marco Velli. This work was started at a  workshop in ISSI, Bern ``Small-scale transient phenomena and their contribution to coronal heating''. NCAR is sponsored by the National Science Foundation. B.D.P., S.W.M., and R.J.L were supported by NASA grants NNX08AH45G, NNX08BA99G, and NNH08CC02C.
\end{acknowledgements}

\bibliographystyle{aa}

\end{document}